\documentclass[aps,prd,preprintnumbers,superscriptaddress,nofootinbib,amsmath,twocolumn,amssymb,showpacs]{revtex4}%,twocolumn showpacs
\usepackage{graphicx}% Include figure files
\usepackage{dcolumn}% Align table columns on decimal point
\usepackage{bm}% bold math
\usepackage{amsmath}
\usepackage{color}
\input{colordvi.tex}
\newcommand{\CO}{{\cal O}}
\newcommand{\vecs}[1]{\mbox{\boldmath${#1}$}}

\newcommand*{\mpl}{M_{\rm Pl}}
\begin{document}

\title{Higgs G-inflation}

\author{Kohei Kamada}
\email[Email: ]{kamada"at"resceu.s.u-tokyo.ac.jp}
\affiliation{Department of Physics, Graduate School of Science,
The University of Tokyo, Tokyo 113-0033, Japan}
\affiliation{Research Center for the Early Universe (RESCEU),
Graduate School of Science, The University of Tokyo, Tokyo 113-0033, Japan}

\author{Tsutomu~Kobayashi}
\email[Email: ]{tsutomu"at"resceu.s.u-tokyo.ac.jp}
\affiliation{Research Center for the Early Universe (RESCEU), Graduate School of Science,
The University of Tokyo, Tokyo 113-0033, Japan}

\author{Masahide~Yamaguchi}
\email[Email: ]{gucci"at"phys.titech.ac.jp}
\affiliation{Department of Physics, Tokyo Institute of Technology, Tokyo 152-8551, Japan}

\author{Jun'ichi~Yokoyama}
\email[Email: ]{yokoyama"at"resceu.s.u-tokyo.ac.jp}
\affiliation{Research Center for the Early Universe (RESCEU), Graduate School of Science,
The University of Tokyo, Tokyo 113-0033, Japan}
\affiliation{Institute for the Physics and Mathematics of the Universe (IPMU),
The University of Tokyo, Kashiwa, Chiba, 277-8568, Japan}

\preprint{RESCEU-28/10}
\begin{abstract}
A new class of inflation models within the context of G-inflation is proposed,
in which the standard
model Higgs boson can act as an inflaton thanks to
Galileon-like non-linear derivative interaction.
The generated primordial density perturbation is shown to be consistent with
the present observational data. % and some concrete predictions are given.
We also make a general discussion on potential-driven
G-inflation models, and find a new consistency relation
between the tensor-to-scalar ratio $r$ and the tensor spectral index $n_T$,
$r = -32 \sqrt{6}n_T / 9$, which is crucial in
discriminating the present models from standard
inflation with a canonical kinetic term.
\end{abstract}
\maketitle

\section{introduction}

Primordial inflation \cite{Sato:1980yn, infrev} is now regarded as a
part of the ``standard'' cosmology because it not only solves the flatness
and the horizon problems but also accounts for the origin of primordial
fluctuations \cite{fluctuation}.  To construct a model of inflation, one
usually assumes a scalar field that drives inflation (called an inflaton)
outside the standard model (SM) of particle physics.  This is
because there are no scalar fields in the SM except for the Higgs boson and
it has been found that the SM Higgs boson cannot be responsible for
inflation as long as its
kinetic term is canonical and it is minimally coupled to gravity~\cite{Linde:1983gd}.
The difficulty here lies in the fact that the self interaction of the SM Higgs boson is so
strong that the resultant primordial density fluctuation would be too large to
be consistent with the present observational data~\cite{Komatsu:2010fb}.

To construct inflation models within the SM,
several variants of Higgs-driven inflation
have been proposed so far. They include models with a non-minimal coupling term to gravity
\cite{nonmin} and with a non-minimal coupling of the Higgs kinetic term with
the Einstein tensor \cite{nonder}.\footnote{Inflation
models where Higgs multiplets act as an inflaton in the context of 
supersymmetric extensions of the SM, {\em e.g.}, the
next-to-minimal supersymmetric 
standard model, have also been proposed in~\cite{nmssm}.
In these models, a non-canonical K$\ddot{\rm a}$hler potential for the 
Higgs multiplet is assumed.}
The amplitude of the curvature perturbation is suppressed due to the large
effective Planck scale in the former case,
while in the latter case the same thing is caused by the enhanced
kinetic function which effectively reduce the self coupling of the Higgs boson.

The simplest way to enhance the kinetic energy would be to add a
non-canonical higher order kinetic term. A number of novel inflation
models with non-standard kinetic terms have been proposed, such as
k-inflation \cite{kinflation}, ghost condensate \cite{ghost}, and Dirac-Born-Infeld
inflation \cite{DBI}.
When incorporating higher order kinetic terms
special care must be taken in order to avoid unwanted ghost instabilities.
Since newly introduced degrees of freedom will lead easily to ghosts,
it would be desirable if the scalar field does not give rise to
a new degree of freedom in spite of its higher derivative nature. 
It has recently been shown that special combinations of
higher order kinetic terms in the Lagrangian produce derivatives no
higher than two both in the gravitational and scalar field equations
\cite{G1,G2}.
A scalar field having this property is often called the
Galileon because it possesses a Galilean shift symmetry in
the Minkowski background.
Such a scalar field has
been studied in the context of modified gravity and dark
energy in~\cite{gde}.
Recently, an inflation model dubbed as ``G-inflation''
was proposed~\cite{GI}, in which inflation is driven by a
scalar field with a Galileon-like kinetic term.
In Ref.~\cite{GI}, the
background and perturbation dynamics of G-inflation
were investigated, revealing interesting
features brought by the Galileon term.
For example,
scale-invariant scalar perturbations
can be generated even in the exactly de Sitter background, and
the tensor-to-scalar ratio can take a significantly larger value than in
the standard inflation models, violating the standard consistency
relation. Other aspects of G-inflation have been explored in
Refs.~\cite{Mizuno:2010ag, Tolley} (see also~\cite{Trincherini}).

In this paper, we propose a new Higgs inflation model
by adding a Galileon-like kinetic term to the
standard Higgs Lagrangian. We show that
a self coupling constant of the order of the unity
is compatible with the present
observational data thanks to the kinetic term enhanced by the
Galileon effect. We however
start with a general discussion on G-inflation
driven by the potential term because our potential-driven G-inflation
is not restricted only to the Higgs field. We first give a criterion to
determine which term becomes dominant in the kinetic term.
%, a Galileon term or a canonical one.
Then, the slow-roll parameters and the
slow-roll conditions are concretely given in terms of the potential
$V(\phi)$ and the function characterizing the Galileon term.
We also derive the
expressions for primordial fluctuations in terms of the slow-roll
parameters, and find a new model-independent consistency relation for a
potential-driven G-inflation model, which is quite useful for
discriminating it from the standard inflation model with a canonical
kinetic term.
It turns out, however, that primordial non-Gaussianity of
the curvature fluctuation is not large in potential-driven G-inflation.
Finally, as a
concrete example of a potential-driven G-inflation model, we propose a
Higgs G-inflation model. This model predicts that the scalar spectral
index $n_s \simeq 0.967$ and the tensor-to-scalar ratio $r \simeq 0.14$
for the number of $e$-folds ${\cal N} = 60$, which, together with the new
consistency relation $r = -32 \sqrt{6} n_T / 9$, makes our Higgs
G-inflation model testable in near future.

This paper is organized as follows. In the next section, we make a
general discussion on the potential-driven G-inflation model. In
Sec. III, we apply it to more concrete examples, which have
chaotic-type, new-type, and hybrid type potential forms. In Sec. IV, we
present a new class of inflation model that regards the standard model
Higgs boson as an inflaton in the context of G-inflation. Final section
is devoted to conclusions and discussion.

\section{potential-driven G-inflation}

The general Lagrangian describing lowest-order G-inflation is of the form~\cite{GI}
\begin{gather}
S=\int d^4 x \sqrt{-g} \left[\frac{\mpl^2}{2}R +K(\phi,X)-G(\phi,X)\Box \phi \right],
\end{gather}
where $\mpl$ is the reduced Planck mass, $R$ is the Ricci scalar and 
$X := -\nabla_\mu \phi \nabla^\mu \phi /2$. 
The main focus of the present paper is G-inflation driven by the potential term with
the kinetic term modified by the $G(\phi, X)$ term. We therefore take
the ``standard'' form of the function $K(\phi, X)$,
\begin{eqnarray}
K(\phi,X)=X-V(\phi),
\end{eqnarray}
while for simplicity we assume the
following form of the $G(\phi, X)$ term,
\begin{eqnarray}
G(\phi,X)=-g(\phi)X.
\end{eqnarray}

\subsection{The background dynamics}

Taking the homogeneous and isotropic metric $ds^2=-dt^2+a(t)^2 d\vecs{x}^2$, 
we have the following basic equations governing the background cosmological dynamics:
\begin{eqnarray}
&&3\mpl^2 H^2 = X\left[1-gH\dot\phi\left(6-\alpha\right)\right]+V,
\label{bg-fr}\\
&&\mpl^2\dot H = -X\left[1-gH\dot\phi\left(3+\eta-\alpha\right)\right],
\label{bg-doth}\\
&&H\dot\phi\left[3-\eta-gH\dot\phi\left(9-3\epsilon-6\eta+2\eta\alpha\right)\right]
+(1+2\beta)V' =0,
\nonumber\\\label{bg-eom}
\end{eqnarray}
where the dot represents derivative with respect to $t$
and the prime with respect to $\phi$.
In the above we have defined
\begin{eqnarray}
\epsilon&:=&-\frac{\dot H}{H^2},
\\
\eta&:=&-\frac{\ddot\phi}{H\dot\phi},
\\
\alpha &:=&\frac{g'\dot\phi}{gH},
\\
\beta &:=&\frac{g''X^2}{V'}.
\end{eqnarray}
We assume that all of these quantities are small:
\begin{eqnarray}
\epsilon,\;\; |\eta|,\;\; |\alpha|,\;\; |\beta| \ll 1.
\end{eqnarray}
The condition $|\alpha|\ll 1$ indicates that $g(\phi(t))$ must be
a slowly-varying function of time.
Equations~(\ref{bg-fr}) and~(\ref{bg-doth}) together with these slow-roll conditions
imply
\begin{eqnarray}
X,\; |gH\dot\phi X |\ll V.
\end{eqnarray}
Thus, the energy density is dominated by the potential $V$
under the slow-roll conditions:
\begin{eqnarray}
3\mpl^2 H^2\simeq V.
\end{eqnarray}
The slow-roll equation of motion for the scalar field is given by
\begin{eqnarray}
3H\dot\phi\left(1-3gH\dot\phi\right)+V'\simeq 0.
\end{eqnarray}
One can consider two different limiting cases here.
The case $|gH\dot\phi |\ll 1$ corresponds to standard slow-roll inflation,
while in the opposite limit, $| gH\dot\phi |\gg 1$,
the Galileon effect alters the scalar field dynamics.
We are interested in the latter case.
Since $9H^2\dot\phi^2\simeq V'/g$,
it is required that $V'/g > 0$
in order for this regime to be realized.
The slow-roll equation of motion can be solved for $\dot\phi$ to give
\begin{eqnarray}
\dot\phi\simeq -{\rm sgn}(g)\mpl \left(\frac{V'}{3gV}\right)^{1/2}.\label{dotphi}
\end{eqnarray}
We have fixed the sign of $\dot\phi$ so that
${\rm sgn}(\dot\phi) = -{\rm sgn}(V')$, i.e.,
the scalar field rolls {\em down} the potential.
This seems to be a natural situation for the scalar field dynamics.
As we will see below, ghost instabilities are avoided provided that $g\dot\phi<0$,
and hence only in this branch the Universe can be stable.
From Eq.~(\ref{dotphi}) we see
\begin{eqnarray}
-gH\dot\phi\simeq \frac{1}{3}(gV')^{1/2}.
\end{eqnarray}
Therefore, the condition that the kinetic term coming from $G(\phi, X)$ is much bigger than
the usual linear kinetic term $X$ is equivalent to
\begin{eqnarray}
gV'\gg 1.
\end{eqnarray}

Using the slow-roll equations, one can rewrite
the slow-roll parameters in terms of the potential as
\begin{eqnarray}
\epsilon &\simeq& \epsilon_{{\rm std}}\frac{1}{(gV')^{1/2}},
\label{ep-pot}
\\
\eta &\simeq&\frac{\tilde\eta_{\rm std}}{2}\frac{1}{(gV')^{1/2}}-\epsilon+\frac{\alpha}{2},
\label{et-pot}
\\
\alpha&\simeq&-\mpl^2\frac{g'}{g}\frac{V'}{V}\frac{1}{(gV')^{1/2}},
\\
\beta&\simeq&\frac{\mpl^4}{36}\frac{g''}{g}\left(\frac{V'}{V}\right)^2\frac{1}{gV'},
\end{eqnarray}
where $\epsilon_{\rm std}$ and $\tilde\eta_{\rm std}$ are
the slow-roll parameters conventionally used for standard slow-roll inflation,
\begin{eqnarray}
\epsilon_{\rm std}:=\frac{\mpl^2}{2}\left(\frac{V'}{V}\right)^2,
\quad
\tilde \eta_{\rm std}:=\mpl^2\frac{V''}{V}.
\end{eqnarray}
Equations~(\ref{ep-pot}) and~(\ref{et-pot}) clearly show that
the Galileon term effectively flatten the potential
thanks to the factor $1/(gV')^{1/2}\;(\ll 1)$.
This implies that in the presence of the Galileon-like derivative
interaction slow-roll inflation can take place even if
the potential is rather steep.

For later convenience we define $\tilde \eta:=\tilde\eta_{\rm std}/2(gV')^{1/2}$.
It will be also useful to note that
\begin{eqnarray}
\frac{g\dot\phi^3}{\mpl^2 H}\simeq -\frac{2}{3}\epsilon.\label{rel}
\end{eqnarray}
This means that even if the Galileon dominates the dynamics of slow-roll inflation, 
the standard part of the Lagrangian remains much larger than the Galileon interaction term, 
\begin{equation}
|K(\phi,X)| \simeq V(\phi) \gg |G(\phi,X)\Box \phi |. 
\end{equation}

Let us make a brief comment on the initial condition for the scalar field.
The field may initially be off along the slow-roll trajectory~(\ref{dotphi}).
As long as ${\rm sgn}(\dot\phi)=-{\rm sgn}(g)$,
the field safely approaches the trajectory~(\ref{dotphi}).
If ${\rm sgn}(\dot\phi)=+{\rm sgn}(g)$ initially, the situation is more subtle,
because the solution would approach another branch of the slow-roll attractor
and the field would go on to climb up the potential.
This is what indeed happens if ${\cal F}, {\cal G} <0$ at the initial moment,
signaling ghost instabilities
[see Eqs.~(\ref{defF})--(\ref{calFandcalG}) below].
Note, however, that in Eq.~(\ref{calFandcalG})
${\cal F}$ and ${\cal G}$ are evaluated along the
slow-roll trajectory; it is therefore possible in principle
that ${\rm sgn}(\dot\phi)=+{\rm sgn}(g)$
but still one has ${\cal F}, {\cal G} >0$ at the initial moment.
In this case the solution approaches the healthy branch of the slow-roll attractor.

\subsection{Primordial fluctuations}

Let us investigate the properties of scalar cosmological perturbations
in potential-dominated G-inflation.
The quadratic action for the curvature perturbation in the unitary gauge, ${\cal R}$,
is given by~\cite{GI}
\begin{eqnarray}
S_2=\mpl^2\int d\tau d^3x\,a^2\sigma\left[\frac{1}{c_s^2}(\partial_\tau{\cal R})^2
-(\Vec{\nabla}{\cal R})^2\right],\label{s2}
\end{eqnarray}
where $\tau$ is the conformal time and
\begin{eqnarray}
\sigma&:=&\frac{X{\cal F}}{\mpl^2\left(H-\dot\phi XG_X/\mpl^2\right)^2},
\\
c_s^2&:=&\frac{{\cal F}}{{\cal G}},
\end{eqnarray}
with
\begin{eqnarray}
{\cal F}&:=& K_X+2G_X({\ddot \phi}+2H{\dot \phi})-2\frac{G_X^2}{\mpl^2}X^2 \notag \\
&&+2G_{XX}X{\ddot \phi}-2(G_\phi-XG_{\phi X}) \label{defF},\\
{\cal G}&:=& K_X+2XK_{XX}+6G_XH{\dot \phi}+6 \frac{G_X^2}{\mpl^2}X^2 \notag \\
&&-2(G_\phi+XG_{\phi X})+6G_{XX}HX{\dot \phi}\label{defG}.
\end{eqnarray}
The above expressions are for general $K(\phi, X)$ and $G(\phi, X)$,
but in the present case we simply have
\begin{eqnarray}
{\cal F}\simeq -4 gH\dot\phi,
\quad{\cal G}\simeq -6g H\dot\phi,\label{calFandcalG}
\end{eqnarray}
and hence
\begin{eqnarray}
\sigma\simeq\frac{4}{3}\epsilon,
\quad
c_s^2\simeq\frac{2}{3},
\end{eqnarray}
where we used Eq.~(\ref{rel}).
Note that
$g\dot\phi<0$ is required to ensure the stability against perturbations,
as seen from
Eq.~(\ref{calFandcalG})

Evaluating the power spectrum from the quadratic action~(\ref{s2})
is a standard exercise; we arrive at
\begin{eqnarray}
{\cal P}_{\cal R} &=&\left.\frac{1}{4\pi^2}\frac{H^2}{2\sigma c_s\mpl^2}\right|_{\tau=1/c_sk}
\nonumber\\
&=&\left.\frac{3\sqrt{6}}{64\pi^2}\frac{H^2}{\mpl^2 \epsilon}\right|_{\tau=1/c_sk}. \label{genpr}
\end{eqnarray}
The spectral tilt, $n_s-1=d\,\ln{\cal P}_{\cal R}/d\,\ln k$, can be evaluated as
\begin{eqnarray}
n_s-1 =  -6\epsilon+3\tilde\eta+\frac{\alpha}{2}, \label{gentilt}
\end{eqnarray}
where the relation $\ddot H/H\dot H =-\epsilon-3\eta+\alpha$
was used.

The tensor perturbations are generated
in the same way as in the usual canonical inflation models, and hence
the power spectrum and the spectral index of the primordial gravitational waves are
given by
\begin{equation}
{\cal P}_T =\frac{8}{\mpl^2}\left.\left(\frac{H}{2\pi}\right)^2\right|_{\tau =1/k},\quad
n_T=-2\epsilon. \label{gentensor}
\end{equation}
Thus, we obtain a new, model-independent consistency relation
between the tensor-to-scalar ratio and the tensor spectral index: 
\begin{equation}
r = 16\sigma c_s = -\frac{32\sqrt{6}}{9}n_T. 
\end{equation}

%Therefore, we have a new model-independent consistency relation for 
%potential dominated G-inflation, 
%\begin{equation}
%r=-\frac{32\sqrt{6}}{9}n_T. 
%\end{equation}

\section{Galilean symmetric models}

In this section, we shall clarify slow-roll dynamics of G-inflation for
three representative forms of the potential.  We consider the simplest
case where the Galileon-type kinetic term respects not only the Galilean
shift symmetry in the Minkowski background, but also the shift symmetry
$\phi\to\phi+$const~\cite{Kawasaki:2000yn} during inflation, {\it
i.e.},
\begin{equation}
|g|=\frac{1}{M^3}=\text{const.}, \label{const-g}
\end{equation}
where $M$ is a mass scale.
Here the sign of $g$ should be chosen to coincide with that of $V^{\prime} (\phi)$.
For~(\ref{const-g}) the $G$ term is $Z_2$ odd, and hence
the slow-roll solution~(\ref{dotphi}) can be realized only in
one side of a $Z_2$-symmetric potential.
Note, however, that the following results can be generalized qualitatively to the cases with
more general $g$ having weak dependence on $\phi$, because
$g(\phi)$ may be practically constant for slowly-rolling $\phi$.
Note also that we assume~(\ref{const-g}) only in the inflationary stage;
$g$ may change globally in $\phi$-space and the detailed shape of $g(\phi)$
would play an important role during the reheating stage after inflation.
From the conservative point of view,
reheating will proceed in the same way as in the usual inflation models
by taking $g(\phi)$ such that $g(\phi)\to 0$ around the minimum of the potential $V(\phi)$.
In this section we focus on the dynamics of $\phi$ in the inflationary stage, and
we will come back to the issue of reheating in Sec.~IV.

%Now we consider several models for $V$ and discuss the conditions for potential dominated G-inflation. 

\subsection{Chaotic inflation}\label{sec:chaotic}

First, let us consider the chaotic inflation
model~\cite{chaotic,Kawasaki:2000yn} for which the potential is given by
\begin{equation}
V(\phi)=\frac{\lambda }{n} \phi^n,
\end{equation}
with $n\, (\geq 1)$ being an integer.
We assume that the field is moving in the $\phi > 0$ side and hence $g=+1/M^3>0$.
In this case, the condition $gV'\gg 1$ is equivalent to 
\begin{equation}
\phi \gg \left(\frac{3 M^{3/2}}{\lambda^{1/2}}\right)^{2/(n-1)} =: \phi_G. 
\end{equation}
Since the slow-roll parameters for $\phi\gg\phi_G$ are given by
\begin{eqnarray}
\epsilon = \frac{n^2\mpl^2 M^{3/2}}{2\lambda^{1/2}\phi^{(n+3)/2} },
\quad
{\tilde \eta} = \frac{n-1}{n}\epsilon,
\end{eqnarray}
potential-driven G-inflation proceeds as long as  
\begin{equation}
\phi\gg \left(\frac{n^2}{2}\frac{\mpl^2M^{3/2}}{\lambda^{1/2}}\right)^{2/(n+3)}
=: \phi_{\epsilon G}. 
\end{equation}

If $\phi_G>\phi_{\epsilon G}$,
one can consider the scenario in which
standard chaotic inflation follows
slow-roll G-inflation.
This scenario is possible if
\begin{eqnarray}
M> \frac{n^{(n-1)/3}}{2^{(n-1)/6}3^{(n+3)/6}} \lambda^{1/3} \mpl^{(n-1)/3} =: M_c.  
\label{mc}
\end{eqnarray}

\begin{figure}[tb]
  \begin{center}
    \includegraphics[keepaspectratio=true,height=55mm]{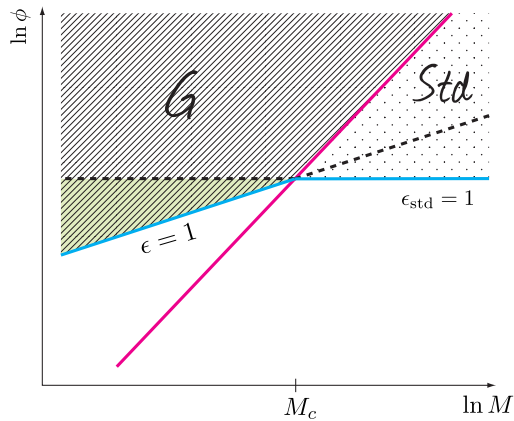}
  \end{center}
  \caption{The Galileon effect operates above the magenta line,
  and the slow-roll condition $\epsilon <1$ is satisfied above the cyan line.
  Chaotic G-inflation therefore takes place in the shaded region, while
  in the dotted region standard chaotic inflation occurs.
  In particular, in the green region the potential is rather steep
  and hence standard inflation would not proceed, but
  G-inflation can. For $M>M_c$ it can be seen that a standard inflationary
  phase follows G-inflation.
  }%
  \label{fig:chaotic.eps}
\end{figure}

If, on the other hand,
$\phi_G < \phi_{\epsilon G}$, {\it i.e.}, $M<M_c$, 
slow-roll G-inflation ends at $\phi=\phi_{\epsilon G}$
and standard chaotic inflation does not follow.
In this case, G-inflation is possible even in the region where
the potential is too steep to support standard chaotic inflation.
In this case, the number of $e$-folds ${\cal N}$ reads
\begin{equation}
{\cal N}=\int_\phi^{\phi_{\epsilon G}} \frac{H}{\dot \phi} d\phi=\frac{2 \lambda^{1/2}}{n (n+3) \mpl^2 M^{3/2}}\phi^{(n+3)/2}-\frac{n}{n+3}.
\end{equation}
From this we obtain
the field value evaluated ${\cal N}$ $e$-folds before the end of inflation,
\begin{equation}
\phi_{\cal N}=\left[
(n+3){\cal N}+n\right]^{2/(n+3)} \left(\frac{n \mpl^2 M^{3/2}}{2 \lambda^{1/2}}\right)^{2/(n+3)}. 
\end{equation}
The situation is summarized in Fig.~\ref{fig:chaotic.eps}.

%%%%%%%%%%%%%%%%%%%%%%%%%
%%%%%%%%%%%%%%%%%%%%%%%%%

Now we investigate the primordial perturbation. 
In the present case we find 
\begin{equation}
{\cal F}\simeq \frac{4}{3}\frac{\lambda^{1/2}\phi^{(n-1)/2}}{M^{3/2}}, \ \ {\cal G}\simeq \frac{2\lambda^{1/2}\phi^{(n-1)/2}}{M^{3/2}}.
\end{equation}
From Eqs. \eqref{genpr} and \eqref{gentilt}, the primordial density perturbation generated 
during the potential-dominated chaotic G-inflation is evaluated as
\begin{align}
{\cal P}_{{\cal R}} \simeq& \frac{\sqrt{6}}{32\pi^2 n^3}\frac{\lambda^{3/2}\phi^{3(n+1)/2}}{M^{3/2}\mpl^{6}} \notag\\
\simeq&  \frac{7.8\times 10^{-3}}{n^3} \notag \\
&\times \left\{\frac{\lambda M^{n}}{\mpl^{4}} \left(\frac{n}{2}
\left[(n+3){\cal N} +n\right]\right)^{(n+1)}\right\}^{3/(n+3)},\\
n_s-1&\simeq -\frac{3(n+1)}{n}\epsilon \simeq -\frac{3(n+1)}{(n+3){\cal N}+n}. 
\end{align}
The scalar-to-tensor ratio is given by
\begin{equation}
r\simeq \frac{64}{3}\left(\frac{2}{3}\right)^{1/2} \epsilon \simeq \frac{17n}{(n+3){\cal N}+n}. 
\end{equation}

For $n=2$ and ${\cal N} \simeq 50$, we obtain $n_s\simeq 0.964$ and $r \simeq 0.13$. 
The COBE/WMAP normalization, ${\cal P}_{{\cal R}} \simeq 2.4 \times 10^{-9}$ 
at $k=0.002 \,{\rm Mpc}^{-1}$~\cite{Komatsu:2010fb}, is attained by taking 
\begin{equation}
\lambda^{1/2}\simeq 3\times 10^{16} \left(\frac{M}{10^{12}\,\,{\rm GeV}}\right)^{-1}\,\,{\rm GeV}. 
\end{equation}
For $n=4$ and ${\cal N} \simeq 60$ we find $n_s\simeq 0.965$ and $r\simeq 0.16$, 
which are also compatible with WMAP~\cite{Komatsu:2010fb}. 
In this case we obtain
\begin{equation}
\lambda \simeq 0.8  \left(\frac{M}{10^{12}\,\,{\rm GeV}}\right)^{-4}, 
\end{equation}
under the COBE/WMAP normalization~\cite{Komatsu:2010fb}, showing that
$\lambda$ can easily be $\sim
\CO(0.1)$.
This
motivates us to study Higgs G-inflation, which will be discussed in the
later section.

\subsection{New inflation}

Next, let us consider the new inflation model \cite{new} where the potential
is given by
\begin{equation}
V(\phi)=V_0-\frac{1}{2}m^2 \phi^2.
\end{equation}
with $V_0 \gg m^2\phi^2/2$. 
Since we consider the range $\phi>0$ and $V'<0$ there, we take $g= - 1/M^3<0$. 
In this case, the condition $g V' \gg1 $ is equivalently written as 
\begin{equation}
\phi \gg \frac{M^3}{m^2} =: \phi_G. \label{phiG}
\end{equation}
The slow-roll parameters can be expressed as
\begin{eqnarray}
\epsilon =
\frac{\mpl^2 m^3M^{3/2}\phi^{3/2}}{2V_0^2},\quad
\tilde\eta = -\frac{\mpl^2mM^{3/2}}{2V_0\phi^{1/2}}.
\label{slpnew}
\end{eqnarray}
Both of the above quantities are smaller than unity
in the range
\begin{eqnarray}
\phi_{\eta G}<\phi<\phi_{\epsilon G},
\end{eqnarray}
where
\begin{eqnarray}
\phi_{\epsilon G}:= \frac{V_0^{4/3}}{\mpl^{4/3}m^2M},
\quad
\phi_{\eta G} := \frac{\mpl^4 m^2M^3}{4V_0^2}=\frac{\eta_{\rm std}^2}{4}\phi_G,
\label{phisl}
\end{eqnarray}
with $|\eta_{\rm std}|=\mpl^2m^2/V_0$.
From this we see that potential-driven G-inflation can occur if
\begin{equation}
M< \frac{V_0^{5/6}}{m \mpl^{4/3}} =: M_{\rm sl}. \label{msl}
\end{equation}

Slow-roll inflation ends anyway at
\begin{eqnarray}
\phi \approx \frac{\sqrt{2V_0}}{m}:=\phi_V.
\end{eqnarray}
If $\phi_V<\phi_G$ then the Galileon effect never operates during inflation.
To have a G-inflationary phase we therefore require $\phi_G<\phi_V$, {\it i.e.},
\begin{eqnarray}
M< m^{1/3}V_0^{1/6} =:M_V.\label{mv}
\end{eqnarray}
Slow-roll G-inflation takes place provided that
Eqs.~(\ref{msl}) and~(\ref{mv}) are both satisfied.
Note that
\begin{eqnarray}
M_V\lessgtr M_{\rm sl}
\;\;\Leftrightarrow\;\; |\eta_{\rm std}|\lessgtr 1.
\end{eqnarray}

\begin{figure}[tb]
  \begin{center}
    \includegraphics[keepaspectratio=true,height=55mm]{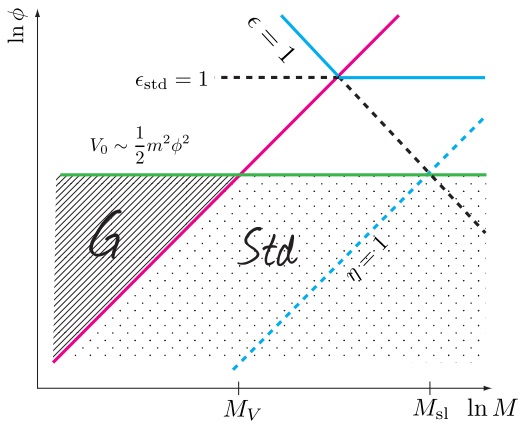}
  \end{center}
  \caption{The same diagram as Fig.~\ref{fig:chaotic.eps}, but
  for new and hybrid inflation with $|\eta_{\rm std}|<1$.
  The Galileon effect operates above the magenta line, the
  slow-roll condition $\epsilon < 1$ is satisfied below the cyan line,
  and the constant piece $V_0$ dominates the potential below the green line.
  }%
  \label{fig:new.eps}
\end{figure}

If $M<M_V<M_{\rm sl}$ ({\it i.e.}, $|\eta_{\rm std}|<1$),
standard new inflation is followed by G-inflation, as shown in Fig.~\ref{fig:new.eps}.
In this case, however, the Galileon term does not help to
support an inflationary phase in the region where
slow-roll inflation would otherwise be impossible.
This case is summarized in Fig.~\ref{fig:new.eps}.

\begin{figure}[tb]
  \begin{center}
    \includegraphics[keepaspectratio=true,height=55mm]{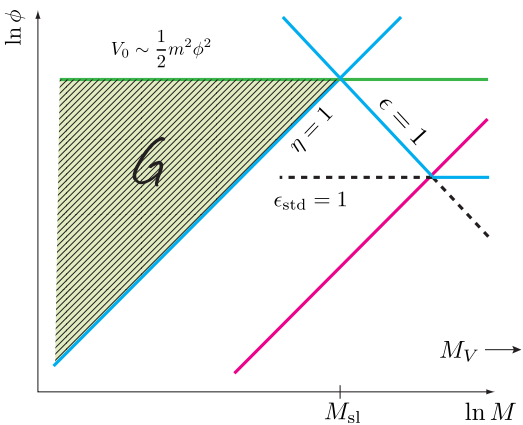}
  \end{center}
  \caption{The same diagram as Fig.~\ref{fig:chaotic.eps}, but for
  new and hybrid inflation with $|\eta_{\rm std}|> 1$. The slow-roll conditions
  are satisfied below the $\epsilon=1$ line and above the $\eta=1$ line. Therefore,
  even though $\eta_{\rm std}>1$, G-inflation can occur in the green, shaded region.
  }%
  \label{fig:new2.eps}
\end{figure}

If $M_V>M_{\rm sl}>M$, $|\eta_{\rm std}|>1$ and hence standard inflation would be impossible.
Nevertheless, slow-roll G-inflation can take place with the help of the Galileon term,
as shown in Fig.~\ref{fig:new2.eps}.

%when 
%the vacuum energy vanishes when 
%\begin{equation}
%\frac{V}{V_0} \ll 1 \Rightarrow  \phi \simeq  \frac{\sqrt{2V_0}}{m} \equiv \phi_V. \label{infendpot}
%\end{equation}
%Note that slow-roll conditions are held where $\phi< \phi_V$. 
%Thus, if
%\begin{equation}
%\phi_G<\phi_V \Leftrightarrow |M|<\frac{2^{1/6}}{3^{2/3}}m^{1/3}V_0^{1/6} \equiv M_V, 
%\end{equation}
%there is a Galileon effect dominated slow-roll solution. 
%As a result, the condition for potential dominated G-inflation is 
%\begin{equation}
%|M| \ll {\rm min.} \{M_{\rm sl}, M_V\}. 
%\end{equation}

%%%%%%%%%%%%%%%%%%%%%%%%%
%%%%%%%%%%%%%%%%%%%%%%%%%

%%%%%%%%%%%%%%%%%%%%%%%%%
%%%%%%%%%%%%%%%%%%%%%%%%%

The number of $e$-folds during new G-inflation is given by
\begin{equation}
{\cal N} = \int_\phi^{\phi_V} \frac{H}{\dot \phi} d\phi\simeq \frac{2V_0}{\mpl^2 M^{3/2}m}
\left(\phi_V^{1/2}-\phi^{1/2}\right). 
\end{equation}
Then, we find the field value, $\phi_{\cal N}$, evaluated
${\cal N}$ $e$-folds before the end of inflation as, 
\begin{equation}
\phi_{\cal N}=\left[\frac{(2V_0)^{1/4}}{m^{1/2}}-\frac{\mpl^2 M^{3/2}m}{2V_0}{\cal N}\right]^2. 
\end{equation}
Except for the special case with $M\sim M_{\rm sl}$
we may have $\phi_V \gg \phi_{\eta G}{\cal N}^2$.
If this is satisfied then we find
\begin{equation}
\phi_{\cal N} \simeq \phi_V. 
\end{equation}

Now we investigate the primordial perturbation. 
In this case, we have
\begin{equation}
{\cal F}\simeq \frac{4m\phi^{1/2}}{3M^{3/2}},  \ \ \ {\cal G} \simeq \frac{2m\phi^{1/2}}{M^{3/2}}. 
\end{equation}
From Eq.~(\ref{genpr}) the power spectrum of the primordial density perturbation generated 
during the potential-dominated new G-inflation is given by
\begin{align}
{\cal P}_{{\cal R}} &\simeq \frac{\sqrt{6}}{32\pi^2}\frac{V_0^3}{m^3 \mpl^6 M^{3/2}\phi^{3/2}}.
\end{align}
More concretely, one can evaluate as
\begin{eqnarray}
{\cal P}_{{\cal R}} &\simeq& 4.6 \times 10^{-3} \frac{V_0^{9/4}}{m^{3/2}\mpl^6 M^{3/2}},
\\
n_s-1 &\simeq&  -6 \epsilon+3 {\tilde \eta} \simeq -6.3 \times \frac{m^{3/2} \mpl^2 M^{3/2}}{ V_0^{5/4}},
\\
r &\simeq& \frac{64}{3}\left(\frac{2}{3}\right)^{1/2}
\epsilon \simeq 14\times \frac{\mpl^2 m^{3/2}M^{3/2}}{V_0^{5/4}}. 
\end{eqnarray}

\subsection{Hybrid inflation}

Finally, let us study the hybrid inflation model \cite{hybrid} where the
potential is effectively given by
\begin{equation}
V(\phi)=V_0+\frac{1}{2}m^2 \phi^2. 
\end{equation}
We consider the range $\phi>0$, and hence take $g=+1/M^3>0$.
Hybrid inflation ends when the waterfall field becomes tachyonic.
Let $\phi_{\rm tac}$ be the value of $\phi$ where this occurs.
We will therefore focus on the range $\phi_{\rm tac} <\phi<\sqrt{2V_0}/m$.
For $\phi\gg \sqrt{2V_0}/m$, the constant piece $V_0$ in the potential can be ignored,
and hence the situation reduces to chaotic inflation studied in Sec.~\ref{sec:chaotic}.

The situation here is analogous to the case of new inflation.
The Galileon effect operates for $\phi\gg\phi_G$, where $\phi_G$
is given in Eq.~(\ref{phiG}).
The slow-roll parameters for $\phi>\phi_{G}$ are given by
\begin{eqnarray}
\epsilon =
\frac{\mpl^2 m^3M^{3/2}\phi^{3/2}}{2V_0^2},\quad
\tilde\eta = \frac{\mpl^2mM^{3/2}}{2V_0\phi^{1/2}},
\end{eqnarray}
so that the slow-roll conditions are satisfied in the range
$\phi_{\eta G}<\phi<\phi_{\epsilon G}$, where $\phi_{\eta G}$ and $\phi_{\epsilon G}$
are the quantities defined in Eq.~(\ref{phisl}).
Thus, the inflaton dynamics can be summarized in Figs.~\ref{fig:new.eps} and~\ref{fig:new2.eps},
depending on the values of $M_V$ and $M_{\rm sl}$.
Since $\dot\phi<0$,
G-inflation is followed by standard hybrid inflation if $M<M_V<M_{\rm sl}$.
If $M<M_{\rm sl}<M_V$ then G-inflation can occur even though
standard inflation would be impossible as $\tilde\eta_{\rm std}>1$.
We remark that hybrid inflation ends at $\phi=\phi_{\rm tac}$, which
is not indicated explicitly in the figures,
and for $\phi>\sqrt{2V_0}/m$ hybrid inflation reduces to chaotic inflation,
which is not shown in the figures either.

The primordial perturbation is described in the same way as in new
inflation,
\begin{align}
{\cal P}_{{\cal R}} &\simeq \frac{\sqrt{6}}{32\pi^2}\frac{V_0^3}{m^3 \mpl^6 M^{3/2}\phi^{3/2}}.
\end{align}

The number of $e$-folds during hybrid G-inflation ${\cal N}$ reads 
\begin{equation}
{\cal N}= \int_\phi^{\phi_{\rm tac}} \frac{H}{\dot \phi} d\phi \simeq \frac{2 V_0}{\mpl^2 M^{3/2} m}
\left(\phi^{1/2}-\phi_{\rm tac}^{1/2}\right). 
\end{equation}
We then obtain the field value evaluated ${\cal N}$ $e$-folds before the end of inflation as, 
\begin{equation}
\phi_{\cal N}=\left(\phi_{\rm tac}^{1/2}+\frac{\mpl^2 M^{3/2}m}{2V_0} {\cal N}\right)^2. 
\end{equation}
In the case where 
\begin{equation}
\phi_{\rm tac} \gg \left(\frac{\mpl^2 M^{3/2}m}{2V_0} {\cal N}\right)^2, 
\end{equation}
we have 
\begin{equation}
\phi_{\cal N} \simeq \phi_{\rm tac},
\end{equation}
and thus
\begin{align}
{\cal P}_{{\cal R}} &\simeq 7.6 \times 10^{-3}
\frac{V_0^3}{m^3 \mpl^6 M^{3/2}\phi_{\rm tac}^{3/2}}, \\
n_s-1 &\simeq 3{\tilde \eta}\simeq \frac{3\mpl^2 m M^{3/2}}{2 V_0 \phi_{\rm tac}^{1/2}}, \\
r & \simeq \frac{64}{3}\left(\frac{2}{3}\right)^{1/2} \epsilon \notag \\
&\simeq 8.7\times \frac{\mpl^2 m^3 M^{3/2} \phi_{\rm tac}^{3/2}}{V_0^2}. 
\end{align}

In the opposite case where 
\begin{equation}
\phi_{\rm tac} \ll \left(\frac{\mpl^2 M^{3/2}m}{2V_0} {\cal N}\right)^2, 
\end{equation}
we have 
\begin{equation}
\phi_{\cal N} \simeq \frac{\mpl^4 M^6m^2}{4 V_0^2}{\cal N}^2,
\end{equation}
so that
\begin{align}
{\cal P}_{{\cal R}} &\simeq 6.2  \times 10^{-3}\frac{V_0^6}{\mpl^{12} m^6 M^6 {\cal N}^3}, \\
n_s-1 &\simeq \frac{3}{\cal N}, \\
r &\simeq 1.1 \times\frac{\mpl^8 m^6 M^6}{V_0^5}{\cal N}^3.
\end{align}

\section{Higgs G-inflation}

Let us now construct a Higgs inflation model in the context of
G-inflation.  The tree-level SM Higgs Lagrangian is
\begin{equation}
S_0=\int d^4 x \sqrt{-g} \left[\frac{\mpl^2}{2}R - |D_\mu {\cal H}|^2-\lambda(|{\cal H}|^2-v^2)^2 \right], 
\end{equation}
where $D_\mu$ is the covariant derivative with respect to the SM gauge
symmetry, ${\cal H}$ is the SM Higgs boson, $v$ is the vacuum
expectation value (VEV) of the SM Higgs and $\lambda$ is the self coupling 
constant.  Since we would like to have a chaotic inflation-like
dynamics of the Higgs boson, we consider the case where its neutral
component $\phi := \sqrt{2}|{\cal H}_0|$ is very large compared with
to $v$: $\phi \gg v$. In this situation, we have only to consider a
simpler action,
\begin{equation}
S_0=\int d^4 x \sqrt{-g} \left[\frac{\mpl^2}{2}R - \frac{1}{2}(\partial_\mu \phi)^2-\frac{\lambda}{4}\phi^4\right]. 
\end{equation}
In addition to the above action, we consider a Galileon-type
interaction, which breaks Galilean shift symmetry weakly,
\begin{align}
S_G&=\int d^4 x \sqrt{-g} \left[ -\frac{2{\cal H}^\dagger }{M^4}D_\mu D^\mu  {\cal H} |D_\mu {\cal H}|^2 \right] \notag \\
&\rightarrow \int d^4 x \sqrt{-g} \left[ -\frac{\phi }{2M^4}\Box \phi (\partial_\mu \phi)^2 \right],  
\end{align}
where $M$ is a mass parameter. Here we assume $M>0$.  Note that gauge
fields that couples to $\phi$ receive heavy mass from the field value of
the Higgs boson and hence we can neglect the effect of gauge fields when
we consider the inflationary trajectory.  This setup corresponds to the
case
\begin{equation}
K=X-V(\phi), \ \ V(\phi)=\frac{\lambda}{4}\phi^4,  \ \ G=-\frac{\phi}{M^4}X. 
\end{equation}

The Galileon effect operates provided that $gV'\gg 1$, {\it i.e.},
\begin{eqnarray}
\phi \gg \lambda^{-1/4} M.\label{gdom}
\end{eqnarray}
In this regime the slow-roll parameters are given by
\begin{eqnarray}
\epsilon=\frac{8\mpl^2M^2}{\lambda^{1/2}\phi^4}
=\frac{4}{3}\tilde\eta =-2\alpha,
\end{eqnarray}
and $\beta=0$, so that the slow-roll conditions are satisfied if
\begin{eqnarray}
\phi \gg 2^{3/4}\lambda^{-1/8}\mpl^{1/2}M^{1/2}=:\phi_{\rm end}.\label{higgsslowcond}
\end{eqnarray}
From Eqs.~(\ref{gdom}) and~(\ref{higgsslowcond}) one can define a mass scale,
analogously to Eq.~(\ref{mc}),
\begin{eqnarray}
M_c:=\lambda^{-3/4} \mpl.
\end{eqnarray}
If $M\ll M_c$, Higgs G-inflation proceeds even if standard Higgs inflation
would otherwise be impossible.
One can draw essentially the same diagram as Fig.~\ref{fig:chaotic.eps}
for Higgs-G inflation.

%In this case, from the Friedmann equation, we have, 
%\begin{equation}
%H=\frac{\lambda^{1/2}\phi^2}{2 \sqrt{3}\mpl}, 
%\end{equation}
%and the equation of motion for Galileon effect dominated scalar fields, 
%\begin{equation}
%{\dot \phi}=-\frac{2 \mpl M^2}{\sqrt{3}\phi}. 
%\end{equation}
%Then the condition for Galileon effect domination $|g V^\prime|\gg 1$  reads, 
%\begin{equation}
%\phi \gg \lambda^{-1/2} M. 
%\end{equation}
%On the other hand, the slow-roll parameters are given by
%\begin{align}
%\epsilon &=\frac{8\mpl^2 M^2}{\lambda^{1/2} \phi^4}, \\
%{\tilde \eta}&=\frac{3}{4}\epsilon, \\
%\alpha &=-\frac{1}{2} \epsilon, \\
%\beta&=0. 
%\end{align} 
%And the condition for the potential domination is 
%\begin{equation}
%\gamma = \epsilon^2
%\end{equation}
%In the case where 
%\begin{equation}
%M \ll \lambda^{1/4} \mpl, 
%\end{equation}
%the condition for Galileon equation domination lasts until the slow-roll condition breaks when 
%\begin{equation}
%\phi_{\rm end}=2^{3/4}\lambda^{-1/8}\mpl^{1/2}M^{1/2} . 
%\end{equation}

The number of $e$-folds ${\cal N}$ is given by
\begin{align}
{\cal N}&=\int^{\phi_{\rm end}}_\phi \frac{H}{\dot \phi} d\phi =\frac{1}{16}\frac{\lambda^{1/2}}{\mpl^2 M^2}\phi^4-\frac{1}{2},
\end{align}
from which we obtain
the field value evaluated ${\cal N}$ $e$-folds before the end of inflation as
\begin{equation}
\phi_{\cal N}=(16{\cal N}+8)^{1/4}\lambda^{-1/8} \mpl^{1/2} M^{1/2} . 
\end{equation}
As will be mentioned at the end of this section,
reheating after Higgs G-inflation proceeds in the same way
as in the standard inflationary models, and hence the history of the Universe
after inflation will not be altered. Therefore, we use the value
${\cal N}_{\rm COBE}=60$.

%After the end of inflation, the Higgs boson will start damped oscillation around the VEV. 
%Due to the higher order kinetic term, the behavior of the oscillation may be rather 
%complicated. 
%However, when the field value becomes less than $M$, 
%the contribution of higher order kinetic term becomes subdominant. 
%Thus the epoch of higher order kinetic term domination is not so long and 
%the background evolution of the Universe will be $\rho_{\rm tot}\propto a^{-4}$ after that. 
%Then, we can estimate the number of $e$-folds, which normalizes the COBE amplitude, 
%${\cal N}_{\rm COBE}\simeq  60$. 

Now let us
turn to the primordial perturbation in this model.
Using the slow-roll approximation, we obtain
\begin{align}
{\cal F}&\simeq  \frac{4\lambda^{1/2} \phi^2}{3M^2}, 
&{\cal G}&\simeq \frac{2\lambda^{1/2} \phi^2}{M^2}. 
\end{align}
We thus arrive at
\begin{align}
{\cal P}_{{\cal R}}&=\frac{1}{4\pi^2} \frac{H^4}{{\dot \phi}^2 c_s^2 \sqrt{{\cal F}{\cal G}}} \notag \\
&= \frac{(2{\cal N}_{\rm COBE}+1)^2 \lambda^{1/2}}{8 \pi^2}\left(\frac{3}{8}\right)^{1/2}\left(\frac{M}{\mpl}\right)^2 \notag \\
&\simeq 1.1 \times 10^2 \lambda^{1/2}\left(\frac{M}{\mpl}\right)^2, 
\end{align}
and
\begin{align}
n_s&=1-4\epsilon \simeq 1-\frac{4}{2{\cal N}_{\rm COBE}+1} \simeq 0.967.
\end{align}
According to WMAP observations,
${\cal P}_{{\cal R}}= 2.4\times 10^{-9}$~\cite{Komatsu:2010fb},
and hence
\begin{equation}
M \simeq 4.7 \times 10^{-6}\lambda^{-1/4}\mpl \simeq  10^{13}{\rm GeV}. 
\end{equation}
The scalar-to-tensor ratio is given by 
\begin{equation}
r=\frac{64}{3}\left(\frac{2}{3}\right)^{1/2} \epsilon \simeq 17\times \frac{1}{2{\cal N}_{\rm COBE}+1}.  
\end{equation}
For ${\cal N}_{\rm COBE}=60$ this yields $r \simeq 0.14$, which
is large enough to be 
detected by the forthcoming observation by  
PLANCK \cite{Planck}.

Note that
in the above discussion we have neglected the quantum corrections.
In order
to have the precise relation between the potential of the Higgs field and the
observational signatures such as the tensor-to-scalar ratio $r$,
we must
know/assume the complete theory valid up to the inflationary scale, which is
left for future study.
However, the qualitative argument will not be
changed even if we take into account quantum effects, because it can be
absorbed by the variation of $M$.

\begin{figure}[tb]
  \begin{center}
    \includegraphics[keepaspectratio=true,height=55mm]{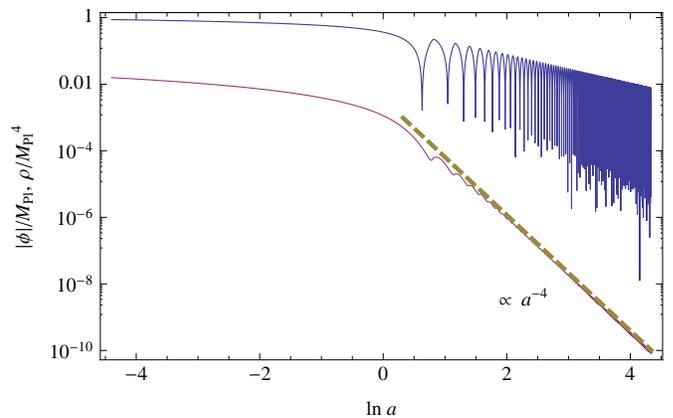}
  \end{center}
  \caption{The evolution of $|\phi|$ (blue, oscillating line) and
  $\rho$ (purple line) in the very final stage of
  Higgs G-inflation and in the reheating stage thereafter.
  We set $a(t_{\rm end})=1$ with $\epsilon(t_{\rm end})=1$.
  The parameters are given by $\lambda=0.1$ and $M=0.01\times M_c$.
  }%
  \label{fig:reheating.eps}
\end{figure}

Before closing this section, let us take a brief look at reheating
after Higgs G-inflation.
The dynamics of the inflaton field during the reheating stage
is non-trivial in general when one considers
non-standard kinetic terms.
In the present case, however, the effect of Galileon-like interaction
can safely be ignored during reheating because $g=\phi/M^4$ is suppressed
around the minimum of the potential, $\phi\ll M$.
In Fig.~\ref{fig:reheating.eps} we show
a numerical example of the
evolution of $\phi$ and $\rho$ in the final stage of Higgs G-inflation
and in the begining of the reheating stage,
when the Higgs field oscillates rapidly.
We have confirmed that the Galileon terms become ineffective
very soon after inflation ends, leading to reheating
in the same way as in the usual case, $\rho\propto a^{-4}$, for the quartic potential.

%%%%%%%%%%%%%%%%%%%%%%%%%
%%%%%%%%%%%%%%%%%%%%%%%%%

\section{Discussion}

The Galileon-like nonlinear derivative interaction, $G(\phi,
X)\Box\phi$, opens up a new arena of inflation model building while
keeping the models healthy, and the novel class of inflation thus
developed ---{\em G-inflation}--- possesses various interesting aspects
to be explored.  In~\cite{GI} the extreme case was emphasized where
G-inflation is driven purely by the kinetic energy, though the
background and the perturbation equations have been derived without
assuming any specific form of $K(\phi,X)$ and $G(\phi,X)$.  In this
paper, we have studied the effects of the Galileon term on
potential-driven inflation.  Although the energy density is dominated by
the potential anyway, the dynamics of the inflaton is nontrivial when
the $G(\phi, X)\Box\phi$ term participates more dominantly than the
usual linear kinetic term $X$.  We have demonstrated that the Galileon
term makes the potential effectively flatter so that slow-roll inflation
proceeds even if the potential is in fact too steep to support
conventional slow-roll inflation.  In light of this fact, we have
constructed a viable model of Higgs inflation, {\it i.e.}, Higgs
G-inflation, showing that the power spectrum of the primordial density
perturbation is compatible with current observational data.  The
tensor-to-scalar ratio $r$ is large enough to be detected by the Planck
satellite.

In this paper we have focused on the power spectrum of the curvature perturbation
and the consistency relation relative to the amplitude of tensor perturbations.
Primordial non-Gaussianity would be another powerful probe to discriminate
G-inflation among others.
Unfortunately, however, non-Gaussianity arising from the present potential-driven models
is estimated to be not large. This is because $f_{\rm NL}$ is composed of
$(1-1/c_s^2)$
and slow-roll suppressed terms with $c_s^2\simeq 2/3$ in the present model,
leading to $f_{\rm NL}\lesssim{\cal O}(1)$.
(Detailed computation of primordial non-Gaussianity from G-inflation
will be given elsewhere~\cite{nong}; see also~\cite{Mizuno:2010ag, Tolley, Trincherini}.)
We would thus conclude that the smoking gun of potential-driven G-inflation
is the consistency relation which is unique enough to distinguish G-inflation
from standard canonical inflation and k-inflation.

We have focused on the (generalized form of the) leading order Galileon term.
As demonstrated in~\cite{gde}, higher order terms play an important role
in the cosmological dynamics of Galileon dark energy models.
Therefore, it would be interesting to consider the effects of the higher order
Galileon terms in the context of primordial inflation, which is left for further study.

\section*{Acknowledgments}

This work was partially supported by JSPS through research fellowships
(K.K.) and the Grant-in-Aid for the Global COE Program ``Global Center
of Excellence for Physical Sciences Frontier''.  This work was also
supported in part by JSPS Grant-in-Aid for Research Activity Start-up
No. 22840011 (T.K.), the Grant-in-Aid for Scientific Research
Nos. 19340054 (J.Y.), 21740187 (M.Y.), and the Grant-in-Aid for
Scientific Research on Innovative Areas No. 21111006 (J.Y.).

\end{document}